\documentclass[useAMS,usenatbib,usegraphicx]{mn2e}

\usepackage{times}

\newcommand{\smallcolsize}{0.40}
\newcommand{\singlecolsize}{0.45}
\newcommand{\middlecolsize}{0.80}
\newcommand{\doublecolsize}{0.90}

\newcommand{\D}{{\rm d}}
\newcommand{\mass}{{\cal M}}
\newcommand{\Msun}{{\rm M}_{\odot}}
\newcommand{\Lsun}{{\rm L}_{\odot}}

\newcommand{\sqdeg}{{\rm deg}^2}

\newcommand{\hunits}{{\rm\,km\,s^{-1}\,Mpc^{-1}}}
\newcommand{\kms}{{\rm\,km\,s^{-1}}}
\newcommand{\effsb}{\mu_{r,50}}
\newcommand{\vmax}{V_{\rm max}}
\newcommand{\vmod}{V'_{\rm max}}

\newcommand{\araa}{ARA\&A}   \newcommand{\aap}{A\&A}
\newcommand{\aj}{AJ}         \newcommand{\apj}{ApJ}
\newcommand{\apjl}{ApJ}      \newcommand{\apjs}{ApJS}
\newcommand{\mnras}{MNRAS}   \newcommand{\nat}{Nature}
     \newcommand{\pasp}{PASP}
\newcommand{\procspie}{Proc.\ SPIE}  \newcommand{\aaps}{A\&AS}

\newcommand{\new}[1]{{#1}} %% do nothing for final version

\title[GAMA: The galaxy stellar mass function]
{Galaxy And Mass Assembly (GAMA): 
The galaxy stellar mass function at $\mathbf{z<0.06}$.}

\author[I.K.~Baldry et al.]
{{\parbox{\textwidth}{\raggedright I.K.~Baldry,$^{1}$
S.P.~Driver,$^{2,3}$
J.~Loveday,$^{4}$
E.N.~Taylor,$^{5,6}$
L.S.~Kelvin,$^{2,3}$
J.~Liske,$^{7}$
P.~Norberg,$^{8}$
A.S.G.~Robotham,$^{2,3}$
S.~Brough,$^{9}$
A.M.~Hopkins,$^{9}$
S.P.~Bamford,$^{10}$
J.A.~Peacock,$^{11}$
J.~Bland-Hawthorn,$^{5}$
C.J.~Conselice,$^{10}$
S.M.~Croom,$^{5}$
D.H.~Jones,$^{12}$
H.R.~Parkinson,$^{11}$
C.C.~Popescu,$^{13}$
M.~Prescott,$^{1}$
R.G.~Sharp,$^{14}$
R.J.~Tuffs$^{15}$
}}\\
\vspace{0.4cm}\\
{\parbox{\textwidth}{\raggedright 
$^{1}$Astrophysics Research Institute, Liverpool John Moores 
University, Twelve Quays House, Egerton Wharf, Birkenhead, CH41 1LD, UK
\\
$^{2}$International Centre for Radio Astronomy Research, University 
of Western Australia, 35 Stirling Hwy, Crawley, WA 6009, Australia
\\
$^{3}$School of Physics \& Astronomy, University of St Andrews, North
Haugh, St Andrews, KY16 9SS, UK
\\
$^{4}$Astronomy Centre, University of Sussex, Falmer, 
Brighton BN1 9QH, UK
\\
$^{5}$Sydney Institute for Astronomy, School of Physics, 
University of Sydney, NSW 2006, Australia
\\
$^{6}$School of Physics, University of Melbourne, Parkville, 
VIC 3010, Australia
\\
$^{7}$European Southern Observatory, Karl-Schwarzschild-Str.~2, 
85748 Garching, Germany
\\
$^{8}$Institute for Computational Cosmology, Department of Physics,
Durham University, South Road, Durham DH1 3LE, UK
\\
$^{9}$Australian Astronomical Observatory, PO Box 296, Epping, 
NSW 1710, Australia
\\
$^{10}$Centre for Astronomy and Particle Theory, University of
Nottingham, University Park, Nottingham NG7 2RD, UK
\\
$^{11}$Institute for Astronomy, University of Edinburgh, Royal
Observatory, Blackford Hill, Edinburgh EH9 3HJ, UK
\\
$^{12}$School of Physics, Monash University, Clayton, VIC 3800, Australia
\\
$^{13}$Jeremiah Horrocks Institute, University of Central Lancashire,
Preston PR1 2HE, UK
\\
$^{14}$Research School of Astronomy \& Astrophysics, The Australian 
National University, Cotter Road, Weston Creek, ACT 2611, Australia
\\
$^{15}$Max Planck Institute for Nuclear Physics (MPIK), 
Saupfercheckweg 1, 69117 Heidelberg, Germany
\\
}}}

\begin{document}

\date{Accepted by MNRAS, 2011 December.}

\pagerange{\pageref{firstpage}--\pageref{lastpage}} \pubyear{2011}

\maketitle

\label{firstpage}

\begin{abstract}
  We determine the low-redshift field galaxy stellar mass function (GSMF)
  using an area of 143 deg$^2$ from the first three years of the Galaxy And
  Mass Assembly (GAMA) survey. The magnitude limits of this redshift survey
  are $r<19.4$\,mag over two thirds and 19.8\,mag over one third of the area.
  The GSMF is determined from a sample of 5210 galaxies using a
  density-corrected maximum volume method. This efficiently overcomes the
  issue of fluctuations in the number density versus redshift. With
  $H_0=70\hunits$, the GSMF is well described between $10^{8}$ and
  $10^{11.5}\Msun$ using a double Schechter function with
%% results 
  $\mass^{*}=10^{10.66}\Msun$,
  $\phi_1^{*}=3.96\times10^{-3}{\rm\,Mpc}^{-3}$, $\alpha_1=-0.35$,
  $\phi_2^{*}=0.79\times10^{-3}{\rm\,Mpc}^{-3}$ and $\alpha_2=-1.47$. 
%% end results
  This result is more robust to uncertainties in the flow-model corrected
  redshifts than from the shallower Sloan Digital Sky Survey main sample
  ($r<17.8$\,mag). The upturn in the GSMF is also seen directly in the
  $i$-band and $K$-band galaxy luminosity functions. Accurately measuring the
  GSMF below $10^{8}\Msun$ is possible within the GAMA survey volume but as
  expected requires deeper imaging data to address the contribution from low
  surface-brightness galaxies.
\end{abstract}


\section{Introduction}
\label{sec:intro}

The galaxy luminosity function (GLF) is a fundamental measurement that
constrains how the Universe's baryonic resources are distributed with galaxy
mass. Before the advent of CCDs and near-IR arrays, the GLF had been primarily
measured in the $B$-band (\citealt*{Felten77,BST88,EEP88};
\citealt{loveday92}).  More recently the low-redshift GLF has been measured
using thousands of galaxies in the redder visible bands
\citep{brown01,blanton03ld} and the near-IR \citep{cole01,kochanek01}, which
more closely follows that of the underlying galaxy stellar mass function
(GSMF).  Furthermore the increased availability of multi-wavelength data and
spectra enables stellar masses of galaxies to be estimated using colours
\citep{LT78,JA92,BdJ01} or spectral fitting (\citealt{kauffmann03A};
\citealt*{PHJ04}; \citealt{gallazzi05}), and either of these methods allow the
GSMF to be computed (\citealt{SP99,balogh01,bell03}; \citealt*{BGD08},
hereafter, BGD08).

Measurement of the GSMF has now become a standard tool to gain insights into
galaxy evolution with considerable effort to extend analyses to $z\sim1$
\citep{drory09,pozzetti10,gilbank11,vulcani11} and higher (\citealt*{EFH08};
\citealt{kajisawa09,marchesini09,caputi11,gonzalez11,mortlock11}).  Overall
the cosmic stellar mass density is observed to grow by a factor of 10 between
$z\sim2$--3 and $z=0$ \citep{dickinson03,EFH08} with significantly less
relative growth in massive $>10^{11}\Msun$ galaxies since $z\sim1$--2
\citep{wake06,pozzetti10,caputi11}.  The evolution in the GSMF is uncertain,
however, with some authors suggesting there could be evolution in the stellar
initial mass function (IMF) \citep{Dave08,wilkins08,vanDokkum08} and
considering the range of uncertainties associated with estimating stellar
masses (\citealt{maraston06}; \citealt*{CGW09}).

The observed GSMF, defined as the number density of galaxies per logarithmic
mass bin, has a declining distribution with mass with a sharp cutoff or break
at high masses often fitted with a \citet{Schechter76} function.  At low
redshift, the characteristic mass of the Schechter break has been determined
to be between $10^{10.6}$ and $10^{11}\Msun$ (\citealt{panter07}; BGD08;
\citealt{LW09}).  The GSMF shape, however, is not well represented by a single
Schechter function with a steepening below $10^{10}\Msun$ giving rise to a
double Schechter function shape overall (BGD08).  \citet{peng10} note that
this shape arises naturally in a model with simple empirical laws for
quenching of star formation in galaxies.  This is one example of the potential
for insights that can be obtained by studying the inferred GSMF as opposed to
comparing observations and theoretical predictions of the GLF; though we note
that it is in some sense more natural for theory to predict the GLF because
model galaxies have a `known' star-formation history.

Abundance matching between a theoretical galactic halo mass function and a GLF
or GSMF demonstrates that, in order to explain the GSMF shape, the fraction of
baryonic mass converted to stars increases with mass to a peak before
decreasing (\citealt{MH02,shankar06}; BGD08; \citealt{CW09,guo10,moster10}).
Galaxy formation theory must explain the preferred mass for star formation
efficiency, the shallow low-mass end slope compared to the halo mass function,
and the exponential cutoff at high masses \citep{oppenheimer10}.  At high
masses, feedback from active galactic nuclei has been invoked to prevent
cooling of gas leading to star formation
\citep{best05,keres05,bower06,croton06}.  The preferred mass scale may
correspond to a halo mass of $\sim 10^{12}\Msun$ above which gas becomes more
readily shock heated \citep{DB06}. Toward lower masses, supernovae feedback
creating galactic winds is thought to play a major role in regulating star
formation \citep{larson74,DS86,LS91,kay02}; while
\citeauthor{oppenheimer10}\ have argued that it is re-accretion of these winds
that is critical in shaping the GSMF.  Others have argued that star formation
in the lowest mass haloes is also suppressed by photoionization
\citep{Efstathiou92,TW96,Somerville02}, in particular, to explain the number
of satellites in the Local Group \citep{benson02}.

Recently, \citet{guo11} used a semi-analytical model applied to the Millenium
Simulation (MS) \citep{springel05nat} and the higher resolution MS-II
\citep{boylan-kolchin09} to predict the cosmic-average `field' GSMF down to
$10^{6}\Msun$. They find that the GSMF continues to rise to low masses reaching
$>0.1$ galaxies per Mpc$^3$ for $10^{6}$--$10^{7}\Msun$, however, they caution
that their model produces a larger passive fraction than is observed amongst
the low-mass population. \citet{mamon11} apply a simple one-equation
prescription, on top of a halo merger tree, that requires star formation to
occur within a minimum mass set by the temperature of the inter-galactic
medium. Their results give a rising baryonic mass function down to
$10^{5.5}\Msun$, with the peak of the mass function of the star-forming galaxy
population at $\sim10^{7}\Msun$. We note that it is useful for theorists to
predict the field GSMF of the star-forming population because measuring this
to low masses, while challenging, is significantly easier than for the passive
population.

Measurements of the GLF reaching low luminosities have been made for the Local
Group \citep{koposov08}, selected groups \citep*{TT02,CKT09}, clusters
\citep{sabatini03,RG08} and superclusters \citep{mercurio06}. To accurately
measure the cosmic-average GSMF, it is necessary to survey random volumes
primarily beyond $\sim10$\,Mpc because: (i) at smaller distances, the
measurement is limited in accuracy by systematic uncertainties in distances to
galaxies \citep*{MHG04}; and (ii) a local volume survey is significantly
biased, e.g., the $B$-band luminosity density out to 5\,Mpc is about a factor
of 5 times the cosmic average (using data from \citealt{karachentsev04} in
comparison with \citealt{norberg02,blanton03ld}).  The Sloan Digital Sky
Survey (SDSS) made a significant breakthrough with redshifts obtained to
$r<17.8$\,mag and multi-colour photometry \citep{stoughton02}, in particular,
with the low-redshift sample described by \citet{blanton05}. In order to
extend and check the low-mass GSMF (BGD08), it is necessary to go deeper over
a still significant area of the sky.

Here we report on the preliminary analysis to determine the $z<0.06$ GSMF from
the Galaxy And Mass Assembly (GAMA) survey, which has obtained redshifts to
$r<19.8$\,mag currently targeted using SDSS imaging but which ultimately will
be updated with deeper imaging.  The plan of the paper is as follows. In
\S~\ref{sec:data}, the data, sample selection and methods are described; in
\S~\ref{sec:results}, the GLF and GSMF results are presented and
discussed. Summary and conclusions are given in \S~\ref{sec:summary}.

Magnitudes are corrected for Galactic extinction using the dusts maps of
\citet*{SFD98}, and are $k$-corrected to compute rest-frame colours and
absolute magnitudes using \textsc{kcorrect} v4\_2 \citep{blanton03kcorr,BR07}.
We assume a flat $\Lambda$CDM cosmology with $H_0=70\hunits$ and
$\Omega_{m,0}=0.3$. The \citet{Chabrier03} IMF (similar to the
\citealt{Kroupa01} IMF) is assumed for stellar mass estimates.  Solar absolute
magnitudes are taken from table~1 of \citet{hill10}, and mass-to-light ratios
are given in solar units.

\section{Data and methods}
\label{sec:data}

\subsection{Galaxy And Mass Assembly survey}

The GAMA survey aims to provide redshifts and multi-wavelength images of
$>250$\,000 galaxies over $>250\,\sqdeg$ to $r<19.8$\,mag
\citep{driver09,driver11,baldry10}. A core component of this programme is a
galaxy redshift survey using the upgraded 2dF instrument AAOmega
\citep{sharp06} on the Anglo-Australian Telescope.  The first three years of
the redshift survey have been completed \citep{driver11} and these data are
used here.  The target selection was over three $48\,\sqdeg$ fields centred at
9\,h (G09), 12\,h (G12) and 14.5\,h (G15) on the celestial equator.  The
limiting magnitudes of the main survey were $r<19.4$ in G09 and G15, $r<19.8$
in G12, $z<18.2$ and $K_{\rm AB}<17.6$ \citep{baldry10}.  It is only the
$r$-band selection that is used in the current analysis because the near-IR
selections add mainly higher-redshift galaxies.  Each area in the survey was
effectively `tiled' 5--10 times with a strategy aiming for high completeness
\citep{robotham10}. In other GAMA papers, \citet{loveday12} determines the
$ugriz$ GLFs, \citet{driver12} determines the cosmic spectral energy
distribution from the far ultraviolet to infrared, while \citet{brough11}
looks at the properties of galaxies at the faint end of the H$\alpha$ GLF.

The target selection was based primarily on SDSS DR6 \citep{sdssDR6} with the
$K$-band selection using UKIRT Infrared Deep Sky Survey (UKIDSS,
\citealt{lawrence07}), and star-galaxy separation using both surveys (see
\citealt{baldry10} for details). Quality control of the imaging selection was
done prior to the redshift survey to remove obvious artifacts and deblended
parts of galaxies. An update to the target list was performed to remove
targets with erroneous selection photometry (\S\,2.9 of
\citealt{driver11}). For this paper, further visual inspection was made of
low-redshift `pairs' with measured velocity differences $<300\kms$.  This
resulted in $\sim50$ objects being reclassified as a deblended part of a
galaxy.  Further inspection was made of targets with faint fibre magnitudes,
reclassifying $\sim100$ objects as not a target. After this, the $r$-band
magnitude limited main survey consists of 114\,360 targets.\footnote{The
  sample was derived from the GAMA database table {\tt TilingCatv16} with
  \textsc{survey\_class} $\ge$ 6.  GAMA \textsc{auto} and Sersic photometry
  were taken from tables {\tt rDefPhotomv01} and {\tt SersicCatv07}; stellar
  masses from {\tt StellarMassesv03}; redshifts, qualities and probabilities
  from {\tt SpecAllv08}; flow-corrected redshifts from {\tt
    DistancesFramesv06}; and photometric redshifts from {\tt Photozv3}.  SDSS
  photometry was taken from SDSS table {\tt dr6.PhotoObj}.}

Various photometric measurements are used in this paper. The selection
magnitudes are SDSS Petrosian magnitudes from the \textsc{photo} pipeline
\citep{stoughton02}. In order to obtain matched aperture photometry from $u$-
to $K$-band, the imaging data from SDSS and UKIDSS were reprocessed and run
through \textsc{SExtractor} \citep{BA96}.  The details are given in
\citet{hill11} and here we use the $r$-defined \textsc{auto} magnitudes
primarily for colours: these use elliptical apertures.  \textsc{SExtractor}
fails to locate some genuine sources that were identified by \textsc{photo},
however, and for these we use Petrosian colours. Finally, an estimate of total
luminosity is obtained using Sersic fits extrapolated to 10$R_e$ (ten times
the half-light radius). This procedure uses a few software packages including
\textsc{galfit} ver.~3 \citep{peng10galfit} and is described in detail by
\citet{kelvin11}.

Spectra for the GAMA survey are taken with the AAOmega spectrograph on the
Anglo-Australian Telescope (AAT), coupled with various other public survey
data and some redshifts from the Liverpool Telescope. The AAT data were
reduced using \textsc{2dFdr} \citep*{CSH04} and the redshifts determined using
\textsc{runz} \citep*{SCS04}.  The recovered redshift for each spectrum is
assigned a quality $Q$ from 1 (no redshift) to 4 (reliable).  These are later
updated based on a comparative analysis between different opinions of a large
subset of spectra. From this process, a best redshift estimate and the
probability ($p_z$) of whether this is correct are assigned to each
spectrum. The new $Q$ values are based on these probabilities (formally called
$nQ$, \S~2.5 of \citealt{driver11}).  Where there is more than one spectrum
for a source, the redshift is taken from the spectrum with the highest $Q$
value.  \new{For the $r$-band limited main sample, 93.1\%, 3.0\% and 3.4\%
  have $Q\ge4$, $Q=3$ and $Q=2$, respectively.} In general, redshifts with
$Q\ge3$ are used, however, $Q=2$ can be considered when there is agreement
with a second spectrum that was measured independently or when there is
reasonable agreement with an independent photometric redshift estimate.

\subsection{Stellar mass estimates}

Stellar masses were computed for GAMA targets using the observed \textsc{auto}
matched aperture photometry \citep{hill11} for the $ugriz$ bands.  These were
fitted using a grid of synthetic spectra with exponentially declining star
formation histories produced using \citet{bc03} models with the
\citet{Chabrier03} IMF\footnote{Stellar masses derived using the Chabrier IMF
  are about 0.6 times the masses derived assuming the Salpeter IMF from
  $0.1\Msun$ to $100\Msun$.} \new{and the \citet{calzetti00} dust obscuration
  law.} The stellar masses were determined from probability weighted integrals
over the full range of possibilities provided by the grid.  See
\citet{taylor11} for details of the method. For the fitted stellar masses in
this paper, we use the stellar mass-to-light ratios (M/L) in $i$-band applied
to the $i$-band Sersic fluxes. Where M/L values are not available \new{(2\% of
  the low-redshift sample)}, we use the colour-based relation of
\citet{taylor11} to estimate M/L$_i$.

The 95\% range in M/L$_i$ from the fitting by \citet{taylor11} is 0.5--2.0 
($\Msun/\Lsun$) for
high luminosity galaxies ($L_i>10^{10}\Lsun$) and 0.2--1.6 for lower
luminosity galaxies ($z<0.06$). The net uncertainty on an individual stellar
mass estimate can be large, e.g., a factor of two or 0.3\,dex as estimated by
\citet{CGW09}. Note though that the impact of uncertainties is more important
when considering evolution in the GSMF than when considering the shape of the
GSMF at a single epoch as in this paper. The latter primarily depends on the
differential systematic uncertainty between populations.
\citet{taylor10masses} estimated that the net differential bias was $\la
0.12{\rm\,dex}$ based on comparing stellar and dynamical mass estimates. The
change in M/L$_i$ between red and blue galaxy populations can be approximated
by a colour-based M/L relation. The effect of changing the slope of this
relation is considered in \S~\ref{sec:gsmf}.

We note that the reason that M/L correlates with colour at all well is that
the M/L of a stellar population increases as a population reddens with age or
dust attenuation. \citet{BdJ01} noted that errors in dust estimates do not
strongly affect stellar mass estimates. In other words, the vectors in M/L
versus dust reddening run nearly parallel to those determined for age
reddening. \citet{driver07}, using the dust models of \citet{popescu00} and
\citet{tuffs04}, confirmed this with the largest deviation for edge-on
systems.

\begin{figure*}
\centerline{ \includegraphics[width=\middlecolsize\textwidth]{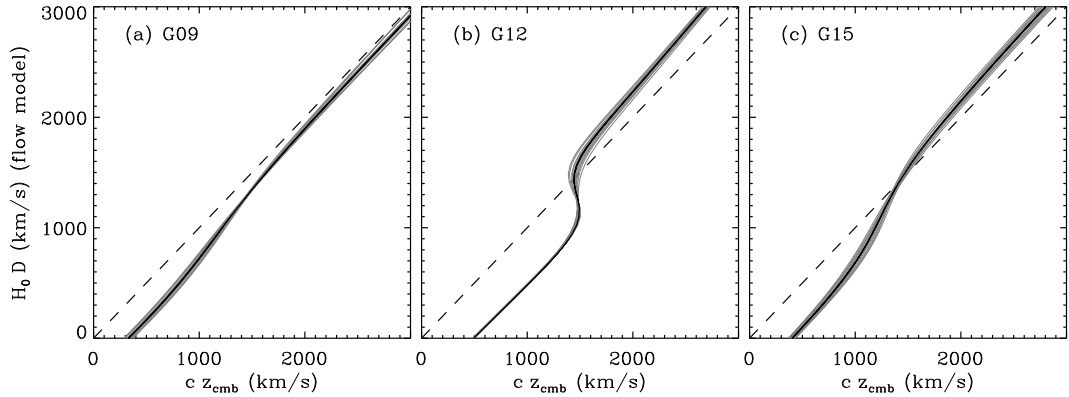} }
\caption{The relation between distance and CMB-frame redshift from the
  \citet{tonry00} flow model. The three GAMA regions are shown in different
  panels. The black line represents the central sight line while the grey
  region shows the variation over each GAMA region.}
\label{fig:flow-model}
\end{figure*}

\subsection{Distances}
\label{sec:distances}

The GAMA survey, as with most large redshift surveys, provides the
heliocentric redshift as standard. In many cases it is sufficient to assume
that this is close to the cosmological redshift. For the GAMA regions at low
redshift, however, it is not.

First we convert heliocentric redshifts to the cosmic microwave background
(CMB) frame:
\begin{equation}
1 + z_{\rm cmb} = (1 + z_{\rm helio}) (1 + z_{\rm sun,comp}) 
\label{eqn:cmb-frame}
\end{equation}
where $c\,z_{\rm sun,comp}$ is the component of the Sun's velocity toward the
object in the CMB frame. We use $z_{\rm sun}=0.001231$ ($369{\rm\,km/s}$) for
the CMB dipole in the direction of ${\rm RA} = 168.0\degr$ and ${\rm
  DEC}=-7.2\degr$ \citep{lineweaver96}.  For the GAMA survey, this leads to
average corrections for the heliocentric velocity ($c\,z_{\rm sun,comp}$) of
$+303$, $+357$ and $+236\kms$, in G09, G12 and G15, respectively.

In the absence of flow information, the CMB frame redshift is a preferred
estimate of the cosmological redshift at $z>0.03$; the velocity of the Local
Group (LG) with respect to the CMB has been attributed to superclusters at
lower redshifts \citep{tonry00,erdogdu06}.  However, it should be noted that
large-scale bulk flows have been claimed by, for example, \citet*{WFH09}.  To
account for flows in the nearby Universe, we use the flow model of
\citeauthor{tonry00}\ linearly tapering to the CMB frame between $z=0.02$ and
$z=0.03$. Figure~\ref{fig:flow-model} shows the relation between the
flow-corrected and CMB frame velocity at $z<0.01$.  The main feature is the
difference between velocities in front of and behind the Virgo Cluster for
sight lines in G12. For each object sky position, the flow-corrected redshift
is obtained by computing the model CMB frame redshifts ($z_{\rm cmb,m}$) for a
vector of cosmological redshifts ($z_{\rm m}$ from $-1000$ to $+1000\kms$ in
steps of 10 around the observed $z_{\rm cmb}$). The flow-corrected redshift
($z$) is then given by the weighted mean of $z_{\rm m}$ values with weights
\begin{equation}
w_m = \exp \left( \frac{-(z_{\rm cmb,m} - z_{\rm cmb})^2}{2 \sigma^2} \right)
\end{equation}
where $\sigma$ is taken as $50\kms$.  This small value is chosen so that the
result is nearly equivalent to using the one-to-one solution for
flow-corrected redshift from $z_{\rm cmb}$, which is mostly available,
\new{and it corresponds to a typical redshift uncertainty from the GAMA
  spectra \citep{driver11}.} It is only around $1500{\rm\,km/s}$ in G12 where
the method is necessary to provide a smoothly varying weighted average between
the three solutions.

\begin{figure}
\centerline{
\includegraphics[width=\smallcolsize\textwidth]{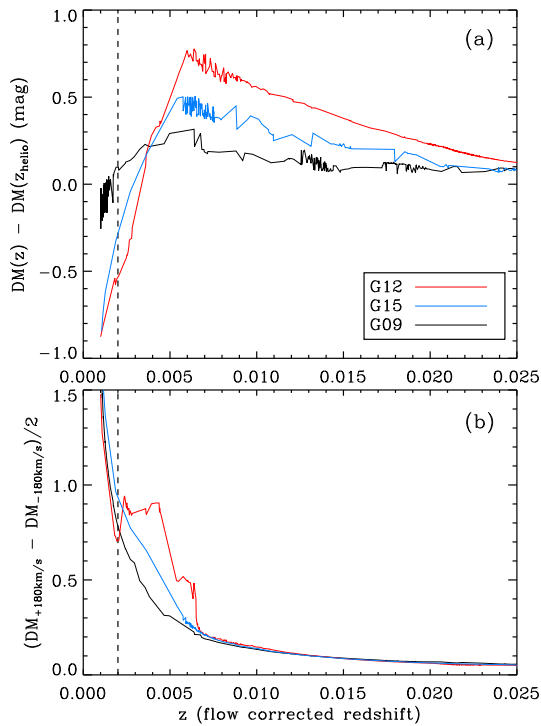}
}
\caption{{\bf (a)}: The difference in the distance moduli derived from $z$ and
  $z_{\rm helio}$. The lines connect the locations of the galaxies for each
  region, which have been sorted by redshift.  The vertical dashed line shows
  the low-redshift limit for the further analyses in this paper.  {\bf (b)}:
  An estimate of the uncertainty in the the distance moduli. The distance
  moduli were recomputed after adjusting the heliocentric redshift by
  $\pm180\kms$ and half this difference is plotted.}
\label{fig:dm-diff}
\end{figure}

Figure~\ref{fig:dm-diff}(a) shows the difference in distance modulus (DM)
using the flow-corrected $z$ compared to using $z_{\rm helio}$ versus
redshift.  Note that the correction to a DM can be larger than 0.5\,mag; the
direction of G12 in particular is within $20\degr$ of both the CMB dipole and
the Virgo Cluster \new{(cf.\ fig.~5 of \citealt{jones06} for the Southern
  sky)}.  The DM uncertainty for each galaxy was estimated by applying changes
in heliocentric velocity of $\pm180{\rm\,km/s}$ and recomputing the
flow-corrected distances.  \new{This corresponds to the cosmic thermal
  velocity dispersion in the \citet{tonry00} model, i.e., the velocity
  deviations after accounting for the attractors.}
Figure~\ref{fig:dm-diff}(b) shows the DM uncertainties, which are taken as
half the difference between the positive and negative changes. The uncertainty
is less than 0.2\,mag at $z>0.007$, while at the lower redshifts the
uncertainty can be quite large especially in G12 because of the triple-valued
solution caused by Virgo infall [Fig.~\ref{fig:flow-model}(b)].

\begin{figure}
\centerline{
\includegraphics[width=\smallcolsize\textwidth]{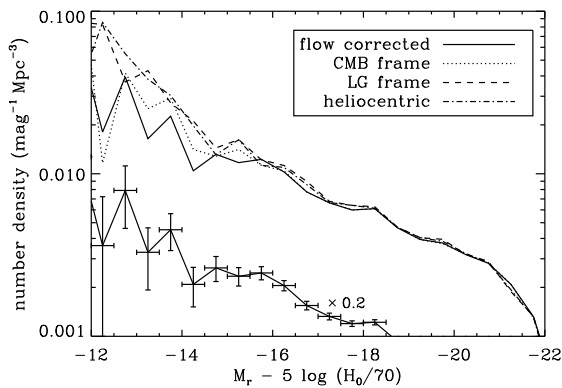}
}
\caption{Comparison between galaxy luminosity functions computed using
  different approximations for the cosmological redshift. The samples were
  selected over the redshift range 0.003 to 0.06, and a standard $\vmax$
  method was used. The solid line with error bars shows the GLF using the
  flow-corrected distances (offset $\times 0.2$, also shown with no offset
  without error bars).}
\label{fig:compare-lf-z}
\end{figure}

To show the significance of using different distances, the $r$-band GLF was
computed using these flow-corrected redshifts, heliocentric frame, LG frame
\citep{CvdB99} and CMB frame \citep{lineweaver96}.
Figure~\ref{fig:compare-lf-z} shows the 4 resultant GLFs.  The computed number
densities are significantly different at magnitudes fainter than $\sim-15$
because of variations in the estimated absolute magnitudes, the sample, and
the volumes $\vmax$. The estimates of number density are lower when using the
flow-corrected redshifts or CMB frame, with respect to heliocentric or LG
frame, because of larger distances and thus higher luminosities with lower
$1/\vmax$ weighting.  \citet{MHG04} noted that the \citet{tonry00} model works
well toward Virgo although possibly at the expense of the anti-Virgo
direction, but in any case, the model is suitable for the GAMA fields.
Hereafter, we use the \citeauthor{tonry00}\ flow-corrected redshifts.

\subsection{Sample selection}
\label{sec:sample-selection}

In addition to the survey selection described in \citet{baldry10}, the
sample selection is as follows:
\begin{enumerate}
\item $r_{\rm Pet} < 19.4$\,mag in G09 or G15, or $r_{\rm Pet} < 19.8$\,mag 
in G12; 
\item redshift quality $Q\ge3$, $Q=2$ when there was agreement with a second
  independent spectrum of the same target (within a velocity difference of
  $450\kms$), or $Q=2$ with $p_z>0.7$ and agreement with a photometric
  redshift estimate [within 0.05 in $\delta z / (1+z)$];
\item $0.002<z<0.06$ (flow corrected), comoving distances from 8.6\,Mpc
  to 253\,Mpc;
\item physical Petrosian half-light radius $>$ 100\,pc. 
\end{enumerate}
\new{The magnitude limits define the $r$-band limited main sample (114\,360)
  with 98.3\% (112\,393) satisfying the redshift quality criteria. The
  redshift range reduces the sample to 5217 (50 of these were included because
  of the $Q=2$ agreement tests). A further 7 sources are rejected by the
  half-light radius criterium giving a primary sample of 5210 galaxies.  The
  100\,pc lower limit corresponds to \citet{gilmore07}'s division between star
  clusters and galaxies. However, on inspection the rejected sources were
  simply assigned an incorrect redshift and should be either stellar or at
  higher redshift than our sample limit.}  

\begin{figure}
\centerline{
\includegraphics[width=\singlecolsize\textwidth]{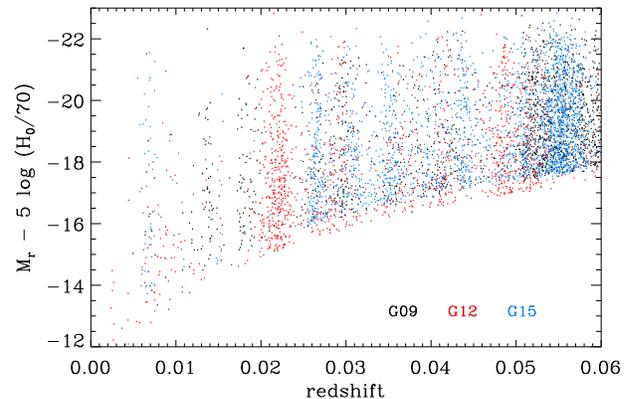}
}
\caption{Absolute magnitude versus redshift.  The black and blue points
  represent galaxies in G09 and G15 to $r<19.4$ while the red points
  represent galaxies in G12 to $r<19.8$.}
\label{fig:z-absmag}
\end{figure}

Figure~\ref{fig:z-absmag} shows the distribution of the primary sample in
$M_r$ versus redshift.  Note not all the redshifts come from the GAMA AAOmega
campaign with the breakdown as follows: 2671 GAMA, 2007 SDSS, 444 2dF Galaxy
Redshift Survey, 64 Millennium Galaxy Catalogue, 10 6dF Galaxy Survey, 6
Updated Zwicky Catalogue, 6 Liverpool Telescope, and 2 others via the
NASA/IPAC Extragalactic Database.

The Petrosian photometry, used for selection of the sample, is highly reliable
having undergone various visual checks. The exception is for overdeblended
sources.  For these the deblended parts have been identified and associated
with a target galaxy. The $r$-band Petrosian photometry of these overdeblended
sources is recomputed by summing the flux from identified parts. About 100
galaxies have their Petrosian magnitude brightened by $>0.1$\,mag from this,
with 14 brightened by more than a magnitude (the target part has not been
assigned the majority of flux in a few cases). It is important to do this
prior to calculating $\vmax$ because a nearby galaxy that is deblended into
parts would not be deblended nearly so significantly if placed at higher
redshift.

\subsection{Density-corrected maximum volume method}
\label{sec:dcvm}

\begin{figure*}
\centerline{
\includegraphics[width=\middlecolsize\textwidth]{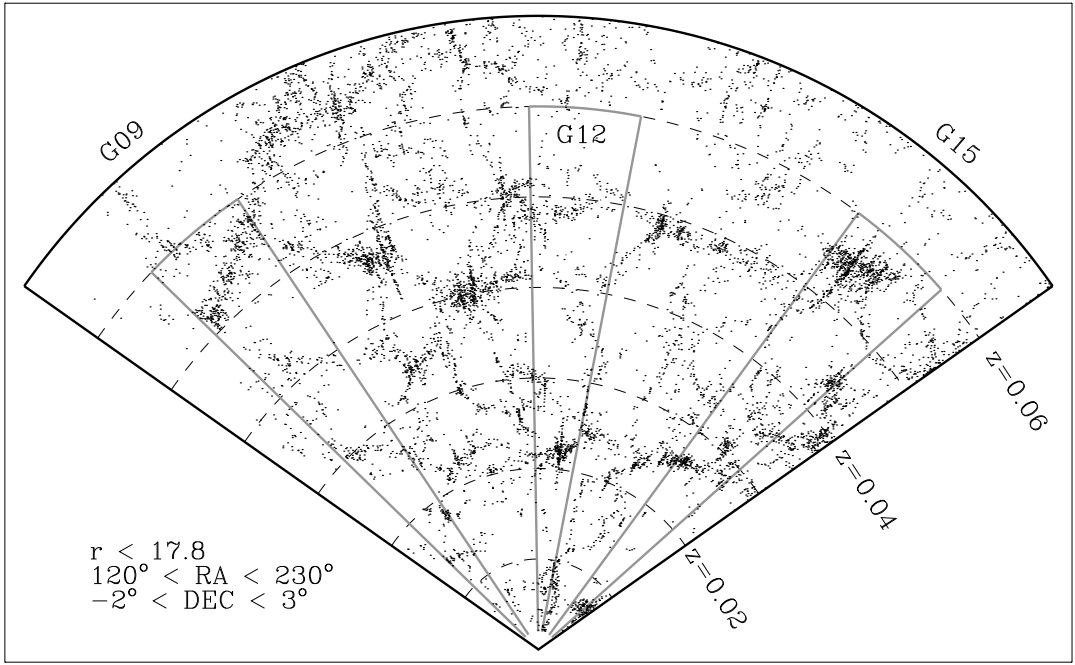}
}
\caption{Redshift distribution in and around the GAMA regions.  Only galaxies
  with $r<17.8$ are shown corresponding to the SDSS main galaxy sample limit
  (but including redshifts from all surveys).  Note the low number of galaxies
  at low RA, to the left of G09 in this figure, is because of the lack of
  redshift survey coverage from SDSS, 2dFGRS and GAMA.}
\label{fig:cone-plot}
\end{figure*}

A standard method to compute binned GLFs is through weighting
each galaxy by $1/\vmax$ \citep{Schmidt68}, which is the comoving volume over
which the galaxy could be observed within the survey limits ($z_{\rm max}$ is
the corresponding maximum redshift).  In the presence of large-scale
structure, large variations in the number density versus redshift, this method
can distort the shape of the GLF
\citep{EEP88}. Figure~\ref{fig:cone-plot} shows the large-scale structure in
and around the GAMA regions.  There are a few substantial overdensities and
underdensities as a function of redshift within each region.

\begin{figure}
\centerline{
\includegraphics[width=\singlecolsize\textwidth]{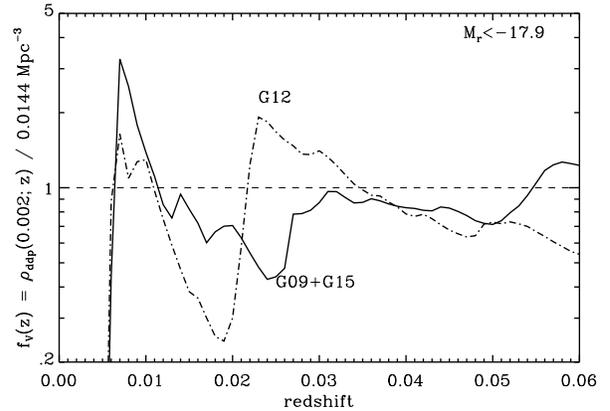}
}
\caption{Variation in number density of the DDP used in the density-corrected
  $\vmod$ method. The lines represent the number density of the $M_r < -17.9$
  volume-limited sample divided by $0.0148{\rm\,Mpc}^{-3}$ from $z=0.002$ up
  to the redshift shown on the $x$-axis.}
\label{fig:zdist-density}
\end{figure}

In order to compute binned GLFs undistorted by radial variations in
large-scale structure, a density-corrected $\vmod$ method is used. This is
given by
\begin{equation}
 \phi_{\log L}  = \frac{1}{\Delta \log L} \,
  \sum_i \frac{1}{V'_{{\rm max},i} \, c_i}
\label{eqn:gsmf-binned}
\end{equation}
where $c_i$ is the completeness factor assigned to a galaxy; and the corrected
volume is given by
\begin{equation}
 V'_{{\rm max},i} = 
  \frac{\rho_{\rm ddp}(z_1; z_{{\rm max},i})}{\rho_{\rm ddp}(z_1; z_2)} \: 
  V_{{\rm max},i} 
\label{eqn:density-correction}
\end{equation}
where $\rho_{\rm ddp}(z_a;z_b)$ is the number density of a density-defining
population (DDP) between redshifts $z_a$ and $z_b$, $z_1=0.002$ is the
low-redshift limit, and $z_2=0.06$ is the high-redshift limit of the sample.
This method is also described in \S~2.7 of \citet{baldry06} where the
density-corrected volume is given by $f_V \vmax$.  To calculate this, we first
treat G09+G15 and G12 separately because of the different magnitude limits,
except that we use a single value for $\rho_{\rm ddp}(z_1; z_2)$, which is
taken to be the average density of the DDP over all three regions. The DDP
must be a volume-limited sample and we use $M_r < -17.9$
(Fig.~\ref{fig:z-absmag}).  Figure~\ref{fig:zdist-density} shows the relative
number density [$f_V(z) = \rho_{\rm ddp}(z_1;z)/\rho_{\rm ddp}(z_1;z_2)$] for
the separate samples.

The redshift upper limit of 0.06 allows sufficient statistics to be obtained
at the bright end to fit the knee of the GLF or GSMF while at the same time
allowing the use of a single DDP that can be used to reliably measure $\vmod$
for galaxies as faint as $M_r \sim -13$. Raising the redshift limit to 0.1
would improve the bright-end statistics at the expense of using a DDP with a
limit that is 1.1\,mag brighter, which is less accurate for determining
$\vmod$ values. We note, however, that it is possible to use a series of
overlapping volume-limited samples to improve the accuracy of $\vmod$
(e.g.\ \citealt{Mahtessian11} `sewed' three samples together).  For the
purposes of keeping a simple transparent assumption and mitigating against
even modest evolution, for this paper we use a single DDP with $z<0.06$.

The step-wise maximum likelihood (SWML) \citep{EEP88} method can also be used
to compute a binned GLF that is not distorted by large-scale structure
variations. In fact, computing the density binned radially and the binned GLF
using a maximum likelihood method can be shown to be equivalent to a
density-corrected $\vmod$ method (\S~8 of \citealt{saunders90};
\citealt{cole11}). This is reassuring but not surprising given that both SWML
and density-corrected $\vmod$ methods assume that the shape of the GLF remains
the same between different regions. This is not exactly true but the resulting
GLF is a weighted radial average. This is seen more transparently in the
density-corrected $\vmod$ method. The real advantage here is that $\vmod$ need
only be calculated for each galaxy using the selection $r$-band Petrosian
magnitudes after which the GLF (or GSMF) can be determined straightforwardly
using different photometry.  When calculating the GLF in a different band (or
the GSMF) there is no colour bias in a bin unless a population with a certain
colour is only visible over a reduced range of luminosity (mass) within the
bin. Note also the GAMA DDP sample is highly complete, which means that the
calculation of $\rho_{\rm ddp}$ is robust.

\begin{figure}
\centerline{
\includegraphics[width=\smallcolsize\textwidth]{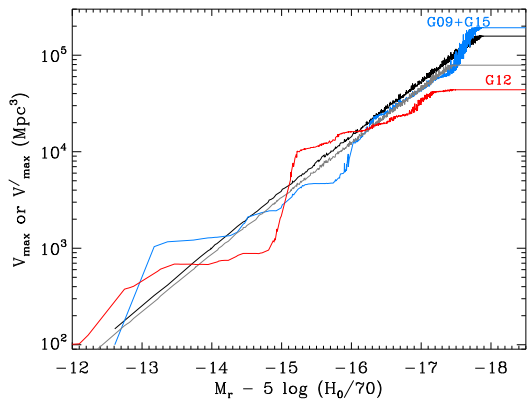}
}
\caption{Comparison between standard maximum volumes and corrected volumes. The
  black and grey lines represent $V_{\rm max}$; note the small scatter is due
  to differential $k$-corrections between $z$ and $z_{\rm max}$. The blue and
  red lines represent $\vmod$. A minimum value of $100{\,\rm Mpc^3}$
  was set for $\vmod$ because this volume could only be expected
  to contain typically 1 or 2 galaxies of the DDP.}
\label{fig:volumes}
\end{figure}

Figure~\ref{fig:volumes} shows a comparison between $\vmod$ and $\vmax$. For
example, note the flattening of $\vmod$ in G12 brighter than $-15.2$ (red
line). This corresponds to the overdensity at $z\simeq0.022$ with the
underdensity beyond.  Brighter galaxies can be seen further but the corrected
volume rises slower than the standard $\vmax$ because the DDP is underdense
beyond.

\begin{figure}
\centerline{
\includegraphics[width=\singlecolsize\textwidth]{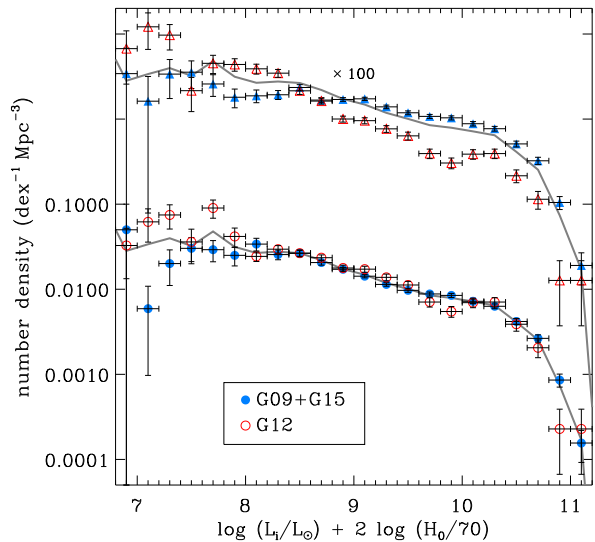}
}
\caption{Comparison between the standard $V_{\rm max}$ (upper data)
  and density-corrected $\vmod$ (lower data) for determining the
  $i$-band GLF.  The triangles show the GLFs from the standard method
  (offset $\times 100$).  Note the major discrepancies between the
  regions. The circles show the GLFs from the $\vmod$ method. The grey
  lines (no offset and offset $\times 100$) represent the GLF using a
  combined $\vmod$ over all three regions.}
\label{fig:compare-methods}
\end{figure}

In order to estimate GLFs, the completeness is assumed to be unity ($c_i=1$)
in this paper with the area of the survey being $143\,\sqdeg$ (one third of
this for each region).  Figure~\ref{fig:compare-methods} shows the $i$-band
GLF computed using the different volume correction methods. The $\vmod$ method
produces much better agreement between the regions than the standard $\vmax$
method. The remaining difference between the regions, below $<10^{8}\Lsun$ in
particular, may be the result of the GLF varying between environments or
uncertainties in the distances. The grey lines in
Fig.~\ref{fig:compare-methods} represent the GLF using a combined volume over
all regions.  This is obtained by modifying $\rho_{\rm ddp}(z_1; z_{{\rm
    max},i}) \, V_{{\rm max},i}$ in Eq.~\ref{eqn:density-correction} to be a
sum over all three regions for each galaxy with $z_{{\rm max},i}$ being
different in G09+G15 ($r<19.4$) compared to G12 ($r<19.8$) (see also
\citealt{AB80} for combining samples with different effective volumes).
Hereafter, this combined $\vmod$ is used. Note also we show GLFs using solar
luminosities because we are working towards the GSMF.

\section{Results and discussion}
\label{sec:results}

\subsection{Galaxy luminosity functions in the $i$-band}

\begin{figure*}
\centerline{
\includegraphics[width=\doublecolsize\textwidth]{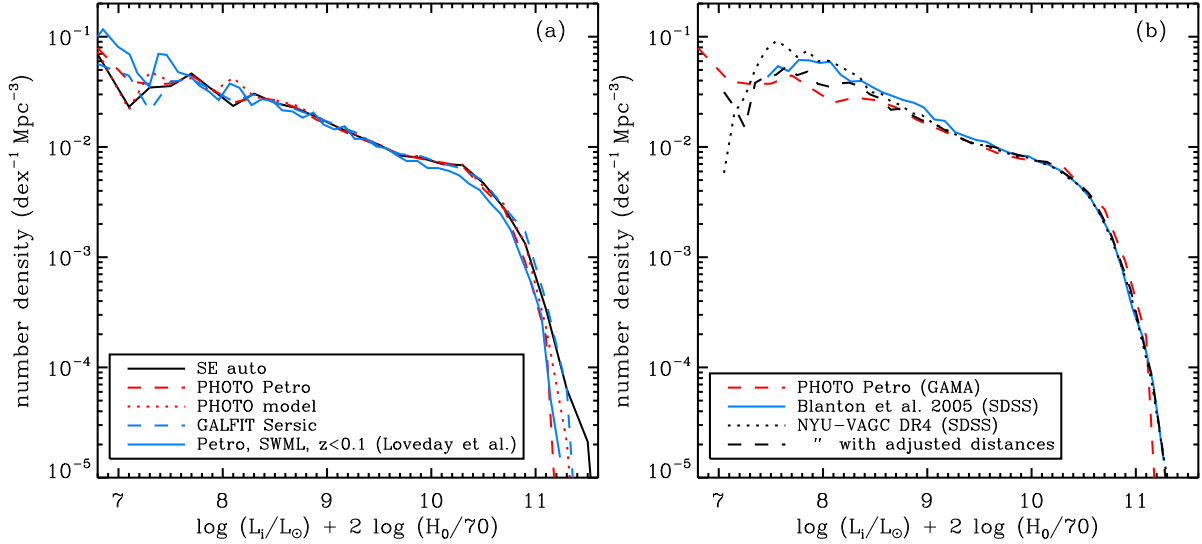}
}
\caption{Comparison between $i$-band GLFs. {\bf (a)}: The red lines, black
  line and blue dashed line represent GLFs computed using the same weights
  ($1/\vmod$) but different aperture photometry from SDSS pipeline
  \textsc{photo}, \textsc{SExtractor} (the \citealt{hill11} GAMA photometry)
  and \textsc{galfit} (the \citealt{kelvin11} Sersic fits). \new{The blue
    solid line shows the result using the SWML method as
    per \citet{loveday12} applied to a $z<0.1$ sample.}  {\bf (b)}: The red
  dashed line is unchanged. The blue line is from the uncorrected (for
  completeness) GLF of \citet{blanton05}, and the black dotted line is
  computed using the same method and sample from BGD08. The black dashed line
  is the same sample with the distances adjusted to the \citet{tonry00} flow
  model.}
\label{fig:compare-lf-magtype}
\end{figure*}

The S/N in the $i$-band is significantly higher than the SDSS $z$-band or any
of the UKIDSS bands for galaxies in our sample. Thus we use the $i$-band as
the fiducial band from which to apply stellar mass-to-light ratios. First we
start by looking at the $i$-band and comparing the GLF taken with different
photometric apertures.  Figure~\ref{fig:compare-lf-magtype}(a) shows the
$i$-band GLF using photometry from (i) SDSS pipeline \textsc{photo}, (ii)
\textsc{SExtractor} as run by \citet{hill11} and (iii) \textsc{galfit} as run
by \citet{kelvin11}.  For comparison, the result from \citet{loveday12} at
$z<0.1$ using Petrosian magnitudes and SWML method is also shown \new{(here
  computed slightly differently to their paper, with $k$-corrections to $z=0$
  and with no completeness corrections).}

The difference between the GLFs in Fig.~\ref{fig:compare-lf-magtype}(a) are
generally small except for the \new{SWML $z<0.1$ GLF which is lower around the
  `knee'.}  The $z<0.1$ GAMA volume is known to be underdense by about 15\%
with respect to a larger SDSS volume (see fig.~20 of \citealt{driver11})
whereas the $z<0.06$ GAMA density is similar to the SDSS volume.  \new{When
  the SWML method is applied to a $z<0.06$ GAMA sample, there is significantly
  better agreement with the density-corrected $\vmod$ LF as expected. Thus the
  $z<0.1$ LF has a different normalisation and shape primarily because the
  $i$-band LF is not exactly universal between different environments.}

The faint end differences in Fig.~\ref{fig:compare-lf-magtype}(a) are
generally not significant (cf.\ error bars in
Fig.~\ref{fig:compare-methods}). At the bright end, the differences are
because the \textsc{auto} apertures and Sersic fits are recovering more flux
from early-type galaxies than the Petrosian aperture.

Figure~\ref{fig:compare-lf-magtype}(b) compares the GAMA result using
\textsc{photo} Petrosian magnitudes with results using the SDSS NYU-VAGC
low-redshift sample ($0.0033 < z < 0.05$; \citealt{blanton05nyuvagc}).
Ignoring the differences below $10^{7.5}\Lsun$, which are because of the
differing magnitude limits, the \citet{blanton05} GLF (DR2) gives a higher
number density below $10^{9}\Lsun$.  This can be at least partly explained by
the distances used.  The NYU-VAGC uses distances from the \citet{willick97}
model, tapering to the LG frame beyond 90\,Mpc.  The black dotted line in
Fig.~\ref{fig:compare-lf-magtype}(b) represents the $i$-band GLF calculated
using the method and sample of BGD08 (DR4) with the NYU-VAGC distances, while
for the black dashed line the distances were changed to those derived from the
\citet{tonry00} model. The latter model gives on average 10\%, and up to 30\%,
larger distances at $z<0.01$.  The DR4 result with the adjusted distances is
in better agreement with the GAMA result. Note that GAMA galaxies with
luminosities $\simeq10^{8}\Lsun$ have a median redshift of 0.02 compared to
0.006 for the NYU-VAGC sample.  Thus the GAMA result is less sensitive to the
flow model at these luminosities.  See \citet{loveday12} for more details on
the GAMA GLFs.

\begin{figure}
\centerline{
\includegraphics[width=\smallcolsize\textwidth]{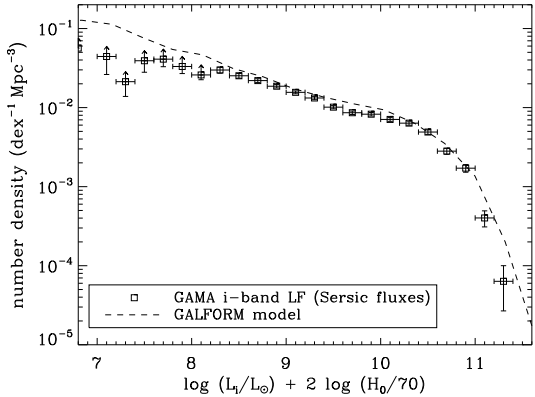}
}
\caption{GAMA data in comparison with a model LF. The model LF was derived by
  H.~Kim \& C.~Baugh using \textsc{galform} \citep{bower06}.}
\label{fig:compare-model}
\end{figure}

Figure~\ref{fig:compare-model} shows the GAMA $i$-band GLF with error bars in
comparison with the GLF from a semi-analytical model. The latter was derived
by H.~Kim \& C.~Baugh (private communication) using an implementation of
\textsc{galform} similar to \citet{bower06}.  The mass resolution of the halo
merger trees was improved and the photo-ionisation prescription was changed so
that cooling in haloes with a circular velocity below $30\kms$ (previously
$50\kms$) is prevented after reionisation ($z=6$). There is reasonable
agreement between the model and data, however, the model LF is higher
particular below $10^{8.2}\Lsun$.  At low luminosities, it is expected that
the GAMA data are incomplete because of surface brightness issues
and the LF data points are shown as lower limits
\new{(justified in the following section, \S~\ref{sec:sb-limit}).}

\subsection{Surface brightness limit}
\label{sec:sb-limit}

In addition to the explicit magnitude limit, there is an implicit and
imprecisely-defined surface brightness (SB) limit that plagues measurements of
the faint end of a GLF \citep{PD86,CD02}. \citet{blanton05} estimated the
impact on the SDSS GLF, determining a completeness of about 0.7 at
$\effsb=23.0{\rm\,mag\,arcsec}^{-2}$ where this is the SB within the Petrosian
half-light radius. Three sources of incompleteness were considered:
photometric incompleteness determined from simulations that put fake galaxies
in frames run through \textsc{photo}, tiling incompleteness because some of
the SDSS area was targeted on versions of \textsc{photo} where the deblender
was not performing optimally, and spectroscopic incompleteness.  The tiling
incompleteness is not an issue here, the issues associated with the
photometric incompleteness may be less severe at the GAMA faint limit because
for a given SB the galaxies are smaller, meaning fewer problems with deblender
shredding and sky subtraction, and the spectroscopic incompleteness can be
mitigated by repeated observations of the same target where necessary.

\begin{figure}
\centerline{
\includegraphics[width=\smallcolsize\textwidth]{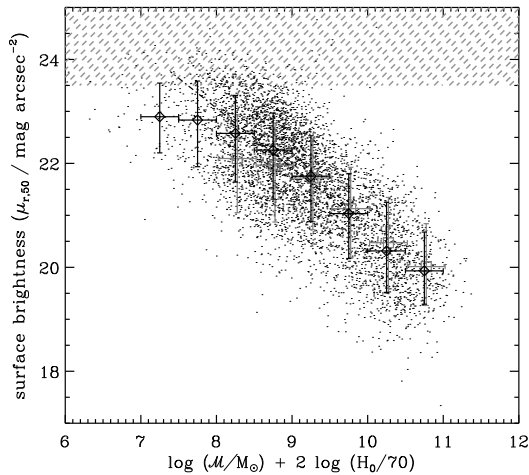}
}
\caption{Surface brightness versus stellar mass. The black diamonds and bars
  represent the GAMA sample used in this paper, while the grey squares and
  bars show the results from an SDSS $z<0.05$ sample (as per fig.~4 of BGD08).
  The vertical bars represent the measured scatter, $\pm1\sigma$, around the
  median in bins of 0.5\,dex. $\effsb$ is the SB within the Petrosian
  half-light radius.  The grey dashed-line region represents the expected area
  of low completeness. The dash-and-dotted line shows the blue population
  relation converted from \citet{geller11}.}
\label{fig:mass-sb}
\end{figure}

Figure~\ref{fig:mass-sb} shows the SB versus stellar mass distribution (with
masses from the colour-based M/L relation of \citealt{taylor11}, see
\S~\ref{sec:gsmf}).  It is difficult to determine when the input catalogue
becomes incomplete.  Judging from the slightly higher mean SB at
$10^{8}$--$10^{9}\Msun$ in the GAMA sample compared to SDSS, we expect that
incompleteness becomes significant for surface brightnesses slightly fainter
than the \citet{blanton05} estimate.  Recently \citet{geller11} analysed a
$0.02<z<0.1$ sample from the Smithsonian Hectospec Lensing Survey
($R<20.6$). They determined a linear relation between SB and $M_R$ for the
blue population.  This is shown in Fig.~\ref{fig:mass-sb} after converting to
$H_0=70$, assuming $\effsb = {\rm SB}_{R,50} + 0.3$ and M/L$_R=0.5$, which is
an average value for a star-forming low-mass galaxy. The average GAMA SB-mass
relation falls below this relation at $\mass < 10^{8}\Msun$, which is where we
expect incompleteness to become significant.  Rather than attempting to
correct for this incompleteness, we assume that the GSMF values below
$10^{8}\Msun$ are lower limits. \new{For the $i$-band LF, values below
  $10^{8.2}\Lsun$ were taken to be lower limits (Fig.~\ref{fig:compare-model})
  because the M/L$_i$ of dwarf galaxies around $10^{8}\Msun$ is typically less
  than 0.8 from the fitting of \citet{taylor11}.}

\subsection{Galaxy stellar mass functions}
\label{sec:gsmf}

Various authors have suggested that M/L in the $i$-band or $z$-band correlates
most usefully with $g-i$ (\citealt{GB09}; \citealt*{ZCR09};
\citealt{taylor10}).  The parametrization is usually linear as follows
\begin{equation}
\log ( \mass / L_i ) = a + b (g-i)
\label{eqn:color-based-ml}
\end{equation}
where $\mass$ is the stellar mass and $L_i$ is the luminosity in solar
units. However, estimates of $a$ and $b$ can vary considerably.
\citet{bell03} give $a=-0.152$ and $b=0.518$ while reading off from figure~4
of \citet{ZCR09} gives $a\simeq-1$ and $b\simeq1$, though the latter is for
resolved parts of galaxies. From fitting to the GAMA data, \citet{taylor11}
obtained $a=-0.68$ and $b=0.73$. This is close to the values obtained from
fitting to SDSS colours and the \citet{kauffmann03A} stellar mass estimates,
for example.

\begin{figure*}
\centerline{
\includegraphics[width=\doublecolsize\textwidth]{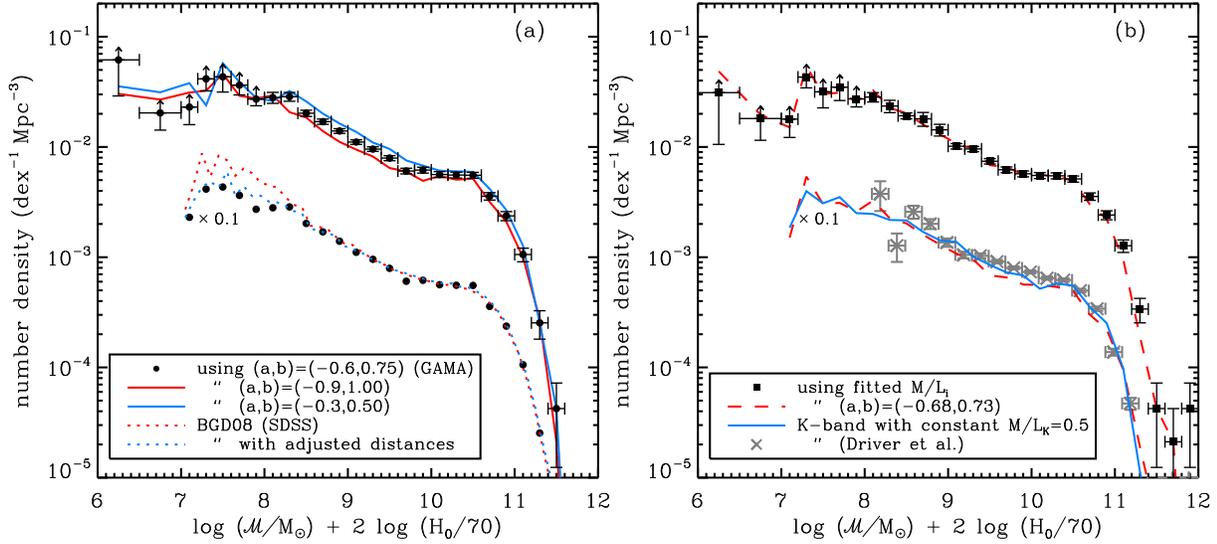}
}
\caption{Galaxy stellar mass functions. {\bf (a)}: These GSMFs were derived
  from Petrosian magnitudes.  The symbols with error bars represent the GAMA
  data using a color-based M/L$_i$ (Eq.~\ref{eqn:color-based-ml}); the colours
  were derived from the GAMA \textsc{auto} photometry.  The red and blue solid
  lines were computed using different $a,b$ values.  The red and blue dotted
  lines (offset $\times 0.1$) represent the results from BGD08 and the
  recomputation using the same sample after changing the flow model to that
  used here, respectively.  {\bf (b)}: The symbols use the fitted GAMA stellar
  M/L$_i$ from \citet{taylor11} applied to the $i$-band Sersic fluxes, while
  the dotted line uses a color-based M/L$_i$. The solid line (offset $\times
  0.1$) shows effectively the $K$-band GLF with $K=K_{\rm auto} - i_{\rm auto}
  + i_{\rm Sersic}$ and applying a constant M/L$_K$, while the crosses with
  error bars show the \citet{driver12} $K$-band GLF.}
\label{fig:gsmf}
\end{figure*}

Figure~\ref{fig:gsmf}(a) shows GSMFs from GAMA data testing the effect of
varying colour-based M/L$_i$. The values of $a$ and $b$ are such that the
M/L$_i$ is 2.0 for galaxies with $g-i=1.2$. The GSMFs show the flattening
around and below $10^{10.5}\Msun$, with a steepening below $10^{9.5}\Msun$:
this is more pronounced with $b=1$ than $b=0.5$.  Also shown is a comparison
with the results of BGD08.  There is generally good agreement between the GAMA
and BGD08 results except at $<10^{8.2}\Msun$. This is despite the fact that
the BGD08 results are expected to be less complete in terms of SB. As noted
above, this is because of the distance model used for the redshifts in
BGD08. If the distances are changed to the model used here then there is
good agreement. The GAMA results are more reliable at $\sim10^{8}\Msun$ because
of the minimal dependence on the distance model. For galaxies with $10^{8} <
\mass / \Msun < 10^{8.4}$, 90\% are brighter than $M_r=-16$, the GLF or GSMF
is not significantly affected by uncertainties in the distances
(Fig.~\ref{fig:compare-lf-z}).

Figure~\ref{fig:gsmf}(b) shows the GSMF from the stellar masses of
\citet{taylor11}, strictly the fitted M/L$_i$ ratios in the \textsc{auto}
apertures applied to the flux derived from the Sersic $i$-band fit (binned
GSMF given in Table~\ref{tab:gsmf}), and the GSMF derived using the best-fit
colour-based M/L$_i$.  These are nearly the same suggesting that a
colour-based M/L$_i$ is easily sufficient for determining a GSMF assuming of
course that it is calibrated correctly.  From the GSMF, the total stellar mass
density is $2.3\times10^{8}\Msun{\rm\,Mpc}^{-3}$. This gives an $\Omega_{\rm
  stars}$ value of 0.0017 (relative to the critical density), or about 4\% of
the baryon density, which is on the low-end of the range of estimates
discussed by BGD08.

\begin{table}
\caption{Galaxy stellar mass function. The $\phi$ values for masses lower than
  $10^{8}\Msun$ should be regarded as lower limits (see
  \S~\ref{sec:sb-limit}). The errors quoted are pseudo Poisson errors derived
  from the square root of the sum of weights squared.}
\label{tab:gsmf}
\begin{tabular}{rrrrr} 
\hline
$\log ({\cal M}/\Msun)$ & bin & $\phi/10^{-3}$ & error & number \\ 
mid point & width & ${\rm dex^{-1}\,Mpc^{-3}}$ & & \\ 
\hline
  6.25&  0.50&    31.1&    21.6&   9\\
  6.75&  0.50&    18.1&     6.6&  19\\
  7.10&  0.20&    17.9&     5.7&  18\\
  7.30&  0.20&    43.1&     8.7&  46\\
  7.50&  0.20&    31.6&     9.0&  51\\
  7.70&  0.20&    34.8&     8.4&  88\\
  7.90&  0.20&    27.3&     4.2& 140\\
  8.10&  0.20&    28.3&     2.8& 243\\
  8.30&  0.20&    23.5&     3.0& 282\\
  8.50&  0.20&    19.2&     1.2& 399\\
  8.70&  0.20&    18.0&     2.6& 494\\
  8.90&  0.20&    14.3&     1.7& 505\\
  9.10&  0.20&    10.2&     0.6& 449\\
  9.30&  0.20&    9.59&    0.55& 423\\
  9.50&  0.20&    7.42&    0.41& 340\\
  9.70&  0.20&    6.21&    0.37& 290\\
  9.90&  0.20&    5.71&    0.35& 268\\
 10.10&  0.20&    5.51&    0.34& 260\\
 10.30&  0.20&    5.48&    0.34& 259\\
 10.50&  0.20&    5.12&    0.33& 242\\
 10.70&  0.20&    3.55&    0.27& 168\\
 10.90&  0.20&    2.41&    0.23& 114\\
 11.10&  0.20&    1.27&    0.16&  60\\
 11.30&  0.20&   0.338&   0.085&  16\\
 11.50&  0.20&   0.042&   0.030&   2\\
 11.70&  0.20&   0.021&   0.021&   1\\
 11.90&  0.20&   0.042&   0.030&   2\\
\hline
\end{tabular}
\end{table}

\subsection{Comparison with the $K$-band galaxy luminosity function}

In order to compare with the shape of the GSMF, we also determined the
$K$-band GLF using the same $\vmod$ values.  For this, we used the $K$-band
magnitude defined by $K=K_{\rm auto} - i_{\rm auto} + i_{\rm Sersic}$, where
the \textsc{auto} photometry is from the $r$-defined catalogue
\citep{hill11}. The reason for this definition is that for low-SB galaxies an
aperture is more accurately defined in the SDSS $r$-band (or $i$-band)
compared to the UKIDSS $K$-band.  This $K-i$ colour is added to our fiducial
$i$-band Sersic magnitude in order to be get a robust estimate of total
$K$-band flux. The resulting GLF was simply converted to a GSMF using M/L$_K =
0.5$, which was chosen to give approximate agreement with the GSMF derived
using the \citet{taylor11} stellar masses. The number densities were divided
by an average completeness of 0.93 because of the reduced coverage in the
$K$-band [fig.~3 of \citet{baldry10}]. This scaled GLF is shown by the blue
line in Fig.~\ref{fig:gsmf}(b). Note that strictly the $\vmod$ values should
be recomputed because of the different coverage across the regions but this
should have minimal impact on the shape. We also show the GAMA $K$-band GLF
from \citet{driver12}, which was derived from a different sample ($0.013 <
z < 0.1$, $r_{\rm Pet}<19.4$ and $K_{\rm AB}< 18.1$ with $r$-defined $K_{\rm
  auto}$ magnitudes) with the same M/L$_K$ applied.

The flattening from $10^{10.6}\Msun$ to $\sim 10^{10}\Msun$ and upturn below
these masses shown in the $i$-band derived GSMF is also seen directly in the
$K$-band GLF [Fig.~\ref{fig:gsmf}(b)].  Though in the case of the
\citet{driver12} result (standard $\vmax$) it is less pronounced.  This is
an important confirmation of this upturn since, while there is some variation
in M/L$_K$, the $K$-band GLF is often used as a proxy for the GSMF. Previous
measurements of the $K$-band field GLF had failed to detect this upturn using
2MASS photometry down to $L_K \la 10^{9}\Lsun$ \citep{cole01,kochanek01} or
using UKIDSS with SDSS redshifts \citep*{SLC09}; see fig 14.\ of
\citeauthor{SLC09} for a compilation.  These measurements nominally probe far
enough down the GSMF ($\sim10^{9}\Msun$) that the upturn should have been
noted.  We note that \citet{merluzzi10}'s measurement of the $K$-band GLF in
the $z=0.048$ Shapley Cluster shows an upturn particularly in the
lower-density environments, however, this does rely on statistical background
subtraction.  The explanation for 2MASS-based GLFs missing this could be the
surface brightness limit. However, GAMA and \citeauthor{SLC09}\ both used
UKIDSS photometry. The difference in this case is that GAMA has redone the
near-IR photometry using $r$-band defined matched apertures \citep{hill11},
and the magnitude limit is higher meaning the galaxies are typically further
away (smaller on the sky) making near-IR photometry more reliable.

\subsection{The double Schechter function}

\begin{figure}
\centerline{
\includegraphics[width=\smallcolsize\textwidth]{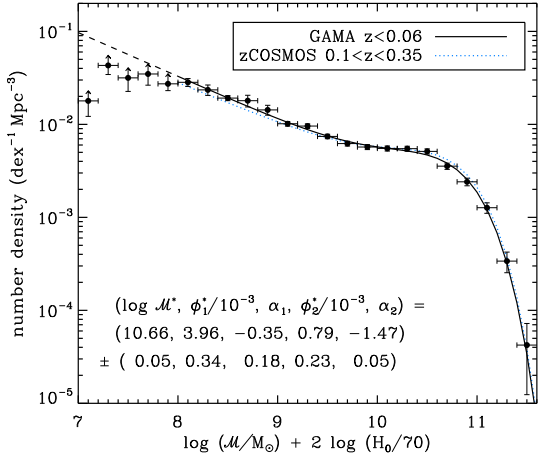}
}
\caption{GSMF with a double Schechter fit to data at $\mass>10^{8}\Msun$. The
  data points represent the GAMA fitted stellar mass results.  The solid line
  represents the fit to the data with extrapolation shown by the dashed line.
  The fit parameters are shown with their 1-sigma errors. Also shown is a fit
  to zCOSMOS data from \citet{pozzetti10}.}
\label{fig:gsmf-fit}
\end{figure}

The shape of the GSMF is well fit with a double Schechter function with a
single value for the break mass ($\mass^{*}$), i.e.\ a five-parameter fit
(BGD08; \citealt{pozzetti10}). This is given by
\begin{equation}
  \phi_\mass \, \D \mass = e^{-\mass/\mass^{*}} \left[ \phi^{*}_1 \left(
  \frac{\mass}{\mass^{*}} \right)^{\alpha_1} + \phi^{*}_2 \left(
  \frac{\mass}{\mass^{*}} \right)^{\alpha_2} \right] \, 
  \frac{ \D \mass }{ \mass^{*} }
\label{eqn:double-schechter}
\end{equation}
where $\phi_\mass \, \D \mass$ is the number density of galaxies with mass
between $\mass$ and $\mass + \D \mass$; with $\alpha_2 < \alpha_1$ so that the
second term dominates at the faintest magnitudes. Figure~\ref{fig:gsmf-fit}
shows this function fitted to the GSMF data providing a good fit. The fit was
obtained using a Levenberg-Marquardt algorithm on the binned GSMF between 8.0
and 11.8 (Table~\ref{tab:gsmf}), and the fit parameters are given in the
plot. The fit to the \citeauthor{pozzetti10}\ GSMF for $z=0.1$--0.35 is also
shown, which is similar.

A natural explanation for this functional form was suggested by
\citet{peng10}. In their phenomenological model, star-forming (SF) galaxies
have a near constant specific star-formation rate (SFR) that is a function of
epoch. Then there are two principle processes that turn SF galaxies into
red-sequence or passive galaxies: `mass quenching' and `environmental
quenching'. In the model, the probability of mass quenching is proportional to
a galaxy's SFR (mass times the specific SFR). This naturally produces a
(single) Schechter form for the GSMF of SF galaxies. Considering only mass
quenching, the GSMF of passive galaxies is also determind to have a Schechter
form with the same value of $\mass^{*}$ but with the faint-end (power-law)
slope $+1$ compared to that of the SF galaxies. To see this consider a single
Schechter function GSMF and multiply by mass: $\mass^{\alpha} \rightarrow
\mass^{\alpha + 1}$. Overall the GSMF of all galaxies is represented by a
double Schechter function with $\alpha_1 = \alpha_2 + 1$.  This is in
agreement with our fit (Fig.~\ref{fig:gsmf-fit}), which has $\alpha_1 -
\alpha_2 = 1.12 \pm 0.19$. In fact, a good fit can be obtained by restricting
$\alpha_1 = \alpha_2 + 1$, making a four-parameter fit (at $\mass > 10^{8}
\Msun$).

In the model, environmental quenching does not change the overall double
Schechter shape as some SF galaxies are turned to red across all masses. The
GSMF of the SF population remains nearly the same shape while the
red-sequence GSMF has a scaled `copy' of the SF GSMF added so that it follows
a double Schechter form most obviously in high density regions.

\begin{figure}
\centerline{
\includegraphics[width=\singlecolsize\textwidth]{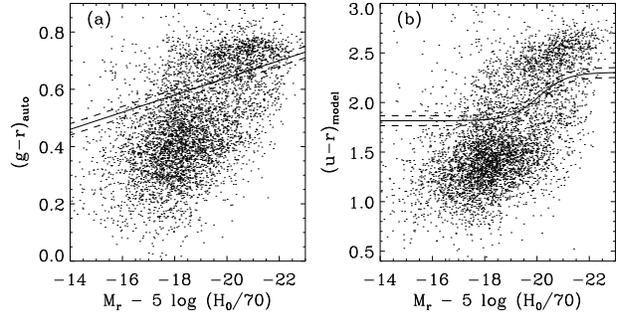}
}
\caption{Color-magnitude diagrams. The points represent the GAMA sample used
  in this paper. The solid and dashed lines represent possible dividing lines
  between the red and blue populations, with slopes of $-0.03$ in (a) and
  using a tanh function from \citet{baldry04} in (b).}
\label{fig:color-divs}
\end{figure}

\begin{figure}
\centerline{
\includegraphics[width=\smallcolsize\textwidth]{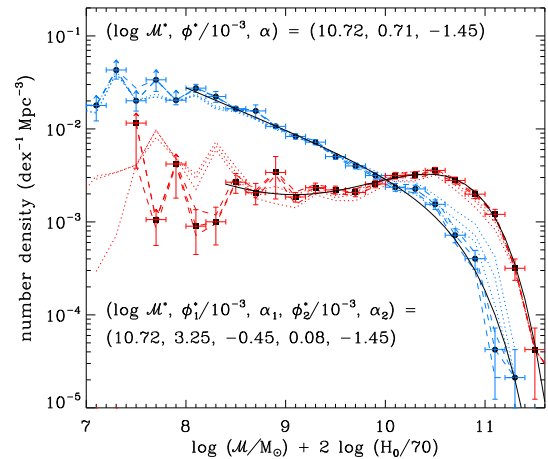}
}
\caption{GSMFs for the red and blue populations. The circles and squares with
  error bars, coloured according to the population, were derived using the
  divider that is shown as a solid line in Fig.~\ref{fig:color-divs}(b). The
  solid lines represent fits to the data.  The dotted and dashed lines
  represent the GSMFs using the six possible dividers based on $g-r$ and
  $u-r$, respectively.}
\label{fig:gsmf-rb}
\end{figure}

To illustrate the origin of the double Schechter shape \new{of the GAMA
GSMF as suggested by} the \citet{peng10} model, 
we divided the galaxies into red and blue populations based on
color-magnitude diagrams. Figure~\ref{fig:color-divs} shows the $g-r$ and
$u-r$ color-magnitude diagrams both versus $M_r$, with three possible dividing
lines using a constant slope of $-0.03$ (e.g.\ \citealt{bell03}) and three
using a tanh function (eq.\ 11 of \citealt{baldry04}), respectively.
Figure~\ref{fig:gsmf-rb} shows the resulting red- and blue-population GSMFs
with the dotted and dashed lines representing the six different colour cuts
(some extremely red objects were not included because the colour measurement
was unrealistic, $g-r>1.0$ or $u-r>3.2$). Following the \citet{peng10} model,
we fit to the red and blue population GSMFs simultaneously with a double
Schechter ($\alpha_1$, $\alpha_2$) and single Schechter function ($\alpha$),
respectively. The fits shown in Fig.~\ref{fig:gsmf-rb} are constrained to
have: the same $\mass^{*}$, $\alpha_2 = \alpha$, and $\alpha_1 = \alpha_2 +
1$. A good fit with the five free parameters is obtained to the two
populations when using a $u-r$ divider. Note there is an excess of blue
population galaxies above a single Schechter fit at high masses when using a
$g-r$ divider, the red population data were not fitted below
$10^{8.4}\Msun$ where there is significant uncertainty in the population type
because of presumably large errors in the colours, and the inclusion of edge-on
dusty disks is a problem for a simple red colour selection. Nevertheless
the basic \citet{peng10} model provides a remarkably simple explanation of the
observed GSMF functional forms.

\subsection{The most numerous type of galaxies}

Are blue (irregulars, late-type spirals) or red (spheroidals, ellipticals)
dwarf galaxies the most numerous type in the Universe (down to
$\sim10^{7}\Msun$)?  Judging from Fig.~\ref{fig:gsmf-rb}, the answer would
appear to be the blue dwarf galaxies, i.e.\ star-forming galaxies.  However,
the measured number densities of both populations may be lower limits; and the
measurement of the red population becomes less reliable below about
$10^{8.4}\Msun$ because of the smaller volume probed, the uncertainties in the
colours, and the cosmic variance is larger because the galaxies are more
clustered than the blue population.

An alternative estimate of the number densities of red galaxies can be
obtained by considering the relative numbers of early-type galaxies in the
Local Group, and then scaling the numbers to match the field GSMF at high
masses ($>10^{9}\Msun$).  This assumes that the Local Group represents an
average environment in which these galaxies are located. Taking the catalogue
of galaxies from \citet{karachentsev04}, galaxies are selected within 1.4\,Mpc
and with Galactic extinction less than 1.2\,mag. The latter excludes two
galaxies viewed near the Galactic plane (a biased direction in terms of
detecting the lowest luminosity galaxies). The $B$-band luminosities are
converted to stellar masses assuming: M/L$_B$ = 3.0 for early-type galaxies
(RC3 type $< 0$); M/L$_B$ = 1.0 for M31, M33 and the Milky Way, which have
already been corrected for internal attenuation; and M/L$_B$ = 0.5 for
late-type galaxies (RC3 type $>6$). From this, there are 6 galaxies with
stellar mass $\ga10^{9}\Msun$, which are M31, M32, M33, M110, the Milky Way
and LMC. For this population the Local Group, taken to cover a volume of
10\,Mpc$^3$, is approximately 50 times higher density than the cosmic average.

\begin{figure}
\centerline{
\includegraphics[width=\smallcolsize\textwidth]{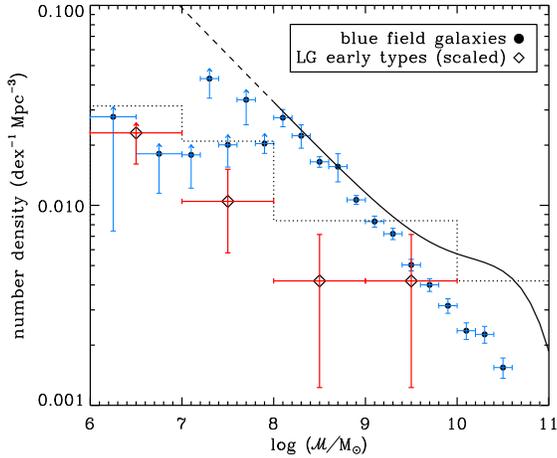}
}
\caption{GSMFs for the field (GAMA) and the Local Group (derived from
  \citealt{karachentsev04}). The solid line shows the GAMA fit from
  Fig.~\ref{fig:gsmf-fit} with the dashed line representing the
  extrapolation. The dotted line represents the scaled LG GSMF in bins of
  1\,dex. The circles represent the blue field population while the diamonds
  represent the LG early types.}
\label{fig:compare-lg}
\end{figure}

Figure~\ref{fig:compare-lg} shows GSMFs for the field and the Local Group
scaled to match, in particular comparing the blue field number densities with
that inferred for the early types by scaling. It is likely that the LG sample
is complete down to $10^7\Msun$ with only some recently discovered satellites
of M31, e.g.\ And XXI \citep{martin09}, suggesting that the bin shown here
from $10^6$ to $10^7$ is a lower limit. This analysis is consistent with the
blue dwarf population being the most common galaxy down to $10^7\Msun$; at
lower masses, it is not yet clear.

\subsection{Future work}

The GAMA GSMF is reliable down to $10^8\Msun$ (corresponding to $M_r \sim -16$
with M/L$_r = 0.5$), which confirms the SDSS result (BGD08) with minor
modification to the distances, assuming that the M/L values are approximately
correct as a function of a galaxy's colour. In addition, there are $\sim350$
galaxies in this GAMA sample between $10^7$ and $10^8\Msun$, and $\sim30$
between $10^6$ and $10^7\Msun$. There are a number of improvements to be made
for the GAMA GSMF measurement at $\mass < 10^8\Msun$:
\begin{enumerate}
\item The GAMA survey is ongoing with an aim to complete redshifts to $r<19.8$
  over 300\,deg$^2$. This will approximately treble the volume surveyed for
  low-luminosity galaxies. 
\item There are about 2000 galaxies so far that have been spectroscopically
  observed twice but with $Q\le2$. A careful coadd of the duplicate
  observations will yield additional redshifts for some of the low-SB
  galaxies.
\item Flux measurements of currently identified low-mass galaxies can be
  improved by careful selection of appropriate apertures.  Automated
  Petrosian or Sersic fitting can lead to large errors for well-resolved
  irregular galaxies.
\item Specialised searches can be made for low-SB galaxies that were missed by
  SDSS \textsc{photo}, in particular, on deeper imaging provided by the
  Kilo-Degree Survey (KIDS) with the VLT Survey Telescope and the VISTA
  Kilo-Degree Infrared Galaxy Survey (VIKING). In the longer term, a
  space-based half-sky survey, such as that planned for the Euclid mission
  \citep{laureijs10} of the European Space Agency, potentially will be able to
  detect low-SB galaxies with $10^5<\mass/\Msun<10^6$ over $10^5{\rm\,Mpc^3}$.
\end{enumerate}

The expected currently missed detection of low-SB galaxies is critical.  In
this sample, the observed number density for galaxies with $10^{6.5}$ to
$10^{7}\Msun$ is only $\sim 0.02\,{\rm Mpc}^{-3}\,{\rm dex}^{-1}$ estimated
using the density-corrected $\vmod$ method. The predicted number by
\citet{guo11} and by extrapolation of the double Schechter function is
$>0.1\,{\rm Mpc}^{-3}\,{\rm dex}^{-1}$. Thus we could be missing significant
numbers of larger low-mass galaxies.

\begin{figure}
\centerline{
\includegraphics[width=\singlecolsize\textwidth]{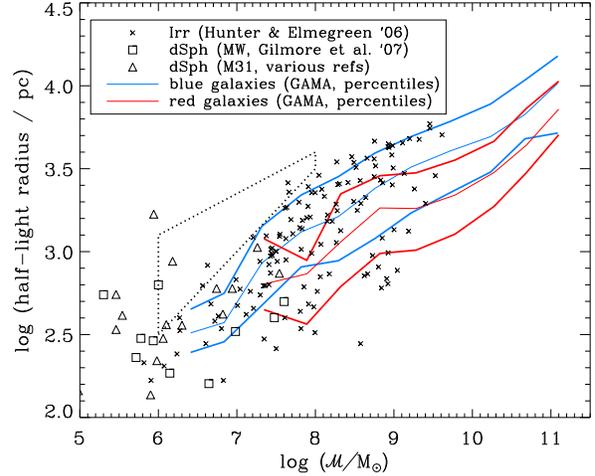}
}
\caption{Size-mass relations. The blue and red solid lines show the 16th,
  median and 84th percentiles of the effective radii from the GAMA Sersic fits
  \citep{kelvin11} for the blue and red populations. The
  symbols represent measurements for irregulars and dwarf spheroidals.  The
  dotted line outlines the region of incompleteness because of SB limits
  relating to the current GAMA sample.}
\label{fig:size-mass}
\end{figure}

Figure~\ref{fig:size-mass} shows the observed size-mass relation of galaxies
from GAMA for blue and red galaxy populations.  For comparison, also shown are
measurements of irregular galaxies \citep{HE06} using M/L$_V$ from the $B-V$
relation of \citet{bell03}, and Milky Way \citep{gilmore07} and M31 dwarf
spheroidals (e.g.\ \citealt{MI06,martin09}) using M/L$_V = 2$.  The GAMA
relation for the blue population follows an approximately linear relation
above $\sim10^{7.5}\Msun$ but appears to drop below the linear extrapolation
at lower masses. The dotted line outlines the region where possible low-SB
galaxies missed by SDSS selection would be located. These would have
$\effsb\sim24$--25\,mag\,arcsec$^{-2}$ for the low M/L blue population.  This
is where an extrapolation of the mass-SB relation to low masses would lie
(Fig.~\ref{fig:mass-sb}). Thus it is essential to use a detection algorithm
that is sensitive to these types of sources (e.g.\ \citealt{kniazev04}) at
distances $\sim$10--50\,Mpc in order to test whether the lowest-mass bins are
incomplete within the GAMA survey volume. For the star-forming population,
obtaining redshifts is feasible but IFUs would be required if only part of
each galaxy has detectable line emission.

The low-redshift sample here only uses 5 per cent of the GAMA $r$-band limited
main survey. The GAMA survey is also well placed to measure the evolution of
the GSMF out to $z\sim0.6$ for the most massive galaxies, study variations
with environment and halo mass, and to study variations in properties with
stellar mass.

\section{Summary and conclusions}
\label{sec:summary}

We present an investigation of the GSMF using the GAMA survey.
\new{Throughout the paper, a recurring theme has been the ways in which
  different aspects of the analysis can affect the inferred shape and
  normalisation of a GLF or GSMF.  In particular we have explored the
  importance of accounting for: bulk flows when estimating distances, large
  scale structure when estimating effective maximum volumes, the effect of
  using different photometric measures, the surface brightness limit, and the
  effect of using different simple prescriptions to estimate stellar mass.}

The distance moduli to apply to the magnitudes depend significantly on using
the \citet{tonry00} flow model in comparison with fixed frames
(Figs.~\ref{fig:flow-model}--\ref{fig:dm-diff}). There is a noticeable effect
on the measured number density of galaxies fainter than $M_r=-16$
(Fig.~\ref{fig:compare-lf-z}). Using the same flow model with SDSS data brings
into better agreement measurements of the GLF and GSMF between SDSS and GAMA
[Fig.~\ref{fig:compare-lf-magtype}(b), Fig.~\ref{fig:gsmf}(a)].  For the same
luminosity galaxies, the $r<19.4$ (19.8 in G12) GAMA sample is less sensitive
to whether the flow model is correct than the $r<17.8$ SDSS sample.

Measuring the GSMF accurately over a large mass range requires surveying a
suitable volume $\ga10^{5}{\rm\,Mpc}^{-3}$ to obtain at least tens of galaxies
at the high-mass end ($\ga10^{11}\Msun$), while the volume over which low-mass
galaxies are observed need not be so large. A problem arises in that the
volume over which a galaxy is visible depends on its luminosity, and any
variations in density as a function of redshift will distort the shape of a
GSMF based on the standard $\vmax$ method. Here we use a density-corrected
$\vmod$ method. This has been shown to be equivalent to a maximum-likelihood
method \citep{cole11} but is simpler to apply to an estimate of the GSMF.  A
volume-limited DDP sample of $M_r<-17.9$ (Fig.~\ref{fig:z-absmag}) was used to
measure relative densities up to a given redshift
(Fig.~\ref{fig:zdist-density}); and these are used to produce the
density-corrected volumes (Eq.~\ref{eqn:density-correction}).  A useful
diagnostic is to plot $\vmod$ versus $M_r$ (Fig.~\ref{fig:volumes}), which
shows that $\vmod$ increases nearly monotonically but with changes in slope
compared to $\vmax$.  The density-corrected $\vmod$ method significantly
reduces the difference in the measured GLFs between the regions compared to
using the standard $\vmax$ method (Fig.~\ref{fig:compare-methods}).

There are small differences in the measured $i$-band GLF depending on the
method of determining a galaxy's flux
[Fig.~\ref{fig:compare-lf-magtype}(a)]. The \textsc{auto} apertures and Sersic
fits recover more flux from early-type galaxies than the Petrosian
aperture. This makes a significant difference at the bright end of the GLF.
Converting the GLF to a GSMF using a colour-based M/L$_i$ relation results in
a more obvious flattening and rise from high to low masses as the $b$
parameter is increased [Fig.~\ref{fig:gsmf}(a)].  Similar GSMF results are
obtained whether using a fitted M/L$_i$ for each galaxy or the colour-based
M/L$_i$ from \citet{taylor11} [Fig.~\ref{fig:gsmf}(b)].  This is not
surprising because the GSMF is only a one-dimensional distribution. We also
find that the $K$-band produces a similar GSMF using a constant M/L$_K=0.5$.
This is an important verification of the upturn based on a simpler assumption
that the $K$-band approximately traces the stellar mass.

As in BGD08 and \citet{pozzetti10}, we find that the double Schechter function
provides a good fit to the data for $\mass>10^{8}\Msun$
(Fig.~\ref{fig:gsmf-fit}). This is approximately the sum of a single Schechter
function for the blue population and double Schechter function for the red
population (Fig.~\ref{fig:gsmf-rb}). This supports the empirical picture,
quenching model, for the origin of the Schechter function by \citet{peng10}.

Blind redshift surveys, like GAMA, are better at characterising the GSMF for
the star-forming field population than the fainter and more clustered red
population. In order to test whether the blue population is the most numerous
in the mass range $10^7-10^8\Msun$ as implied by the GAMA GSMF, we determined
an approximate LG GSMF and scaled the resulting numbers to match the field
GSMF at masses $>10^9\Msun$ (Fig.~\ref{fig:compare-lg}). The numbers of early
types in the cosmic-average GSMF implied by this analysis are below that of
the directly measured blue population.

Accurately measuring the GSMF below $10^8\Msun$ is key to probing new
physics. For example, a simple prescription for preventing star formation in
low-mass haloes, considering temperature-dependent accretion and supernovae
feedback, results in a peak in the GSMF for star-forming galaxies at about
$10^7\Msun$ \citep{mamon11} (note the overall baryonic mass function continues
to rise in their model). The problem with observing low-mass galaxies,
$10^6$--$10^8\Msun$, is not the GAMA spectroscopic survey limit ($r<19.8$) at
least for the star-forming population but primarily the well-known issue with
detecting low-SB galaxies (Fig.~\ref{fig:mass-sb}). Thus a future aim for the
GAMA survey is to characterise the extent of the missing $\sim {\rm kpc}$ size
low-mass population (Fig.~\ref{fig:size-mass}), which ultimately will require
high quality deep imaging with specialised followup.

\section*{Acknowledgements}

Thanks to the anonymous referee for suggested clarifications, and to H.~Kim
and C.~Baugh for providing a model luminosity function.  I.~Baldry and
J.~Loveday acknowledge support from the Science and Technology Facilities
Council (grant numbers ST/H002391/1, ST/F002858/1 and
ST/I000976/1). P.~Norberg acknowledges a Royal Society URF and ERC StG grant
(DEGAS-259586).

GAMA is a joint European-Australasian project based around a spectroscopic
campaign using the Anglo-Australian Telescope. The GAMA input catalogue is
based on data taken from the Sloan Digital Sky Survey and the UKIRT Infrared
Deep Sky Survey. Complementary imaging of the GAMA regions is being obtained
by a number of independent survey programs including GALEX MIS, VST KIDS,
VISTA VIKING, WISE, Herschel-ATLAS, GMRT and ASKAP providing UV to radio
coverage. GAMA is funded by the STFC (UK), the ARC (Australia), the AAO, and
the participating institutions. The GAMA website is
http://www.gama-survey.org/ .

\setlength{\bibhang}{2.0em}
\setlength\labelwidth{0.0em}

%%\bibliographystyle{mn2e-williams}
%%\bibliography{surveys,galaxies,general,stars,two-d-f,fioc-rv}

%%\bsp

\label{lastpage}

\end{document}